\documentclass{article}

\usepackage{mathtools}  % before amsmath
\usepackage{amsmath}
\usepackage{amssymb}  % intercal

\usepackage{spconf}

\usepackage[final]{microtype}
\usepackage{graphicx}

\usepackage{booktabs}  % nicer looking tables
\usepackage[font=small,labelfont=bf]{caption}
\usepackage[hidelinks,pdfpagemode=UseNone]{hyperref}
\usepackage{url}

\usepackage[capitalise]{cleveref}

\usepackage{url}
\usepackage{siunitx}
\DeclareSIUnit{\nounit}{\relax}

\newcommand{\durhm}[2]{\SI{#1}{\hour}\,#2}  % short form duration hours, minutes
\usepackage[%
  olditem,  % do not modify itemize environment
  oldenum   % do not modify enumerate environment
]{paralist}

\DeclareMathOperator{\softmax}{softmax}
\DeclareMathOperator{\Attention}{Attention}

\makeatletter
\newcommand*{\transpose}{%
  {\mathpalette\@transpose{}}%
}
\newcommand*{\@transpose}[2]{%
  \raisebox{\depth}{$\m@th#1\intercal$}%
}
\makeatother

\DeclarePairedDelimiter{\rint}\lfloor\rceil

\newcommand{\mlabel}[1]{\texttt{\mbox{#1}}}

\newcommand\thanksanywhere[1]{%
  \begingroup
  \renewcommand\thefootnote{}\footnote{#1}%
  \addtocounter{footnote}{-1}%
  \endgroup
}
\title{Sequence-to-sequence Singing Synthesis Using the Feed-forward Transformer}

\name{Merlijn Blaauw, Jordi Bonada}
\address{%
Music Technology Group, Universitat Pompeu Fabra, Barcelona, Spain\\
{\small\href{mailto:merlijn.blaauw@upf.edu}{\nolinkurl{merlijn.blaauw@upf.edu}}, \href{mailto:jordi.bonada@upf.edu}{\nolinkurl{jordi.bonada@upf.edu}}}%
}

\begin{document}

\maketitle

\begin{abstract}
We propose a sequence-to-sequence singing synthesizer, which avoids the need for training data with pre-aligned phonetic and acoustic features. Rather than the more common approach of a content-based attention mechanism combined with an autoregressive decoder, we use a different mechanism suitable for feed-forward synthesis. Given that phonetic timings in singing are highly constrained by the musical score, we derive an approximate initial alignment with the help of a simple duration model. Then, using a decoder based on a feed-forward variant of the Transformer model, a series of self-attention and convolutional layers refines the result of the initial alignment to reach the target acoustic features. Advantages of this approach include faster inference and avoiding the exposure bias issues that affect autoregressive models trained by teacher forcing. We evaluate the effectiveness of this model compared to an autoregressive baseline, the importance of self-attention, and the importance of the accuracy of the duration model.%151 words (guidelines: 100--150 word abstract)
\end{abstract}

% 5 keywords (guidelines: upto 5 keywords)

% list widths for preparing figures
%\noindent column width: \SI{\convertto{cm}{\the\columnwidth}}{\centi\meter} \\
%\noindent text width: \SI{\convertto{cm}{\the\textwidth}}{\centi\meter}

%%%%%%%%%%%%%%%%%%%%%%%%%%%%%%%%%%%%%%%%%%%%%%%%%%%%%%%%%%%%%%%%%%%%%%%%%%%%%%%%%%%%%%%%%%
\section{Introduction}\label{sec:intro}
%%%%%%%%%%%%%%%%%%%%%%%%%%%%%%%%%%%%%%%%%%%%%%%%%%%%%%%%%%%%%%%%%%%%%%%%%%%%%%%%%%%%%%%%%%

% why seq2seq?
In recent years, modern TTS systems have largely moved to sequence-to-sequence (Seq2Seq) models, e.g. \cite{WangY2017_Tacotron,ShenJ2018_Tacotron2,PingW2018_DeepVoice3}, where the alignment between the phonetic or orthographic input sequence and the acoustic output sequence is learned during training and inferred during synthesis. One advantage of this approach is that it leads to a more end-to-end system, in the best case avoiding the need for pre-aligned training data, separate phonetic transcription, or a separate duration model at synthesis. For singing synthesis, not requiring pre-aligned training data is particularly attractive, as many existing tools (e.g. forced alignment with a HMM model) do not yield sufficiently accurate results on expressive singing, often requiring manual correction.

% why feed-forward? why transformer?
A common approach for Seq2Seq models in TTS is to use a content-based attention mechanism, e.g. \cite{GehringJ2017_ConvAttention}, sometimes additionally using location-based information, e.g. \cite{GravesA2013_GMMAttention}. As these mechanisms require access to acoustic information at inference, they are normally used in combination with an autoregressive decoder. Recently, some systems have been proposed that use a feed-forward decoder and an alternative attention mechanism that does not rely on access to acoustic information \cite{PengK2019_ParaNet,RenY2019_FastSpeech}. These notably provide faster, parallelizable inference, and are reported to produce more robust alignments with fewer mispronounced, repeated or skipped phonemes.

In the case of singing synthesis, this feed-forward approach is interesting as it avoids the exposure bias problem \cite{RanzatoMA2016_ExposureBias}, caused by the discrepancy between teacher forced training and fully autoregressive inference. This problem can be especially noticeable in long sustained vowels were prediction errors tend to accumulate over time. Additionally, in our experience reaching similar quality results compared to non-Seq2Seq systems can be quite challenging with standard content-based attention mechanisms.

% limit scope
To facilitate evaluation of different systems, we only model timbre in this work, and assume F0 to be given. Although we use ground truth F0 extracted from recordings, it is feasible to predict F0 from the input score with an external model, or possibly predict it jointly. Related to this, we use WORLD vocoder features \cite{MoriseM2016_WORLD} as the output of our system, rather than the more commonly used mel-spectrogram features, as this allows exact control over the synthesized F0. For the best quality results, Seq2Seq systems are typically combined with a neural vocoder, e.g. \cite{TamamoriA2017_WaveNetVocoder,PrengerR2019_WaveGlow,WangX2019_NSFVocoder}, which can work well from both vocoder or mel-spectrogram features. However, in order to get a better idea of the performance of our model on its own, we do not use this approach in the experiments presented here.

% contributions
The principal contributions of this paper are:
\begin{inparaenum}
  \item Propose a singing synthesizer based on the feed-forward Transformer with a practical Seq2Seq mechanism using an external duration model.
  \item Evaluate the quality of this feed-forward model compared to a baseline autoregressive model.
  \item Evaluate the importance of self-attention.
  \item Evaluate the importance of the accuracy of the duration model used.
\end{inparaenum}

%%%%%%%%%%%%%%%%%%%%%%%%%%%%%%%%%%%%%%%%%%%%%%%%%%%%%%%%%%%%%%%%%%%%%%%%%%%%%%%%%%%%%%%%%%
%\section{Methodology}\label{sec:methodology}
\section{Proposed system}\label{sec:proposed_system}
%%%%%%%%%%%%%%%%%%%%%%%%%%%%%%%%%%%%%%%%%%%%%%%%%%%%%%%%%%%%%%%%%%%%%%%%%%%%%%%%%%%%%%%%%%

% brief overview/concept
In singing synthesis, the alignment between the input phonetic sequence and the output acoustic sequence is strongly constrained by the given musical score. This is a notable difference from TTS, which is generally only weakly constrained by the (average) speech rate. Exploiting this fact, we propose to first generate an approximate initial alignment using note timings and a phoneme duration model. Once the input sequence is roughly aligned to the target output timesteps, we assume that the network is able to gradually refine the alignment through a series of transformations, until reaching something close to the target. Note that this approach is quite different from the approach using content-based attention, as here the initial alignment doesn't use any content at all.

% note about importance of accuracy of duration model
An important point here is that we assume that the accuracy of the phoneme duration model is not critical to the end results. We assume that the decoder is powerful enough to be able to recover from errors in the initial alignment, to a certain degree. At the same time, the initial alignment can never hugely deviate from the true alignment, as it is heavily constrained by the note timings. To see if this assumption is correct, we purposely use a very simplistic duration model, based on average phoneme durations computed on a different dataset whose segmentation was corrected by hand. While language dependent, in this case the phoneme duration model is not singer dependent and the values could simply be copied from a table, without the need for any data with phonetic timings.

\subsection{Model architecture}
% inputs/outputs
The input to our system is a musical score, consisting of a sequence of notes. Each note consists of an onset, duration, pitch, and a sequence of phonemes, typically corresponding to a syllable. In this work, we define the note onset as the vowel onset, and note end as the onset of the following vowel or silence. Additionally, we provide an external F0 to our system, in order to capture the affect of pitch on timbre. The output of our system is sequence of harmonic and aperiodic vocoder features, which in this case are simply concatenated.

\begin{figure*}
  \centering
  \includegraphics[width=17.8cm]{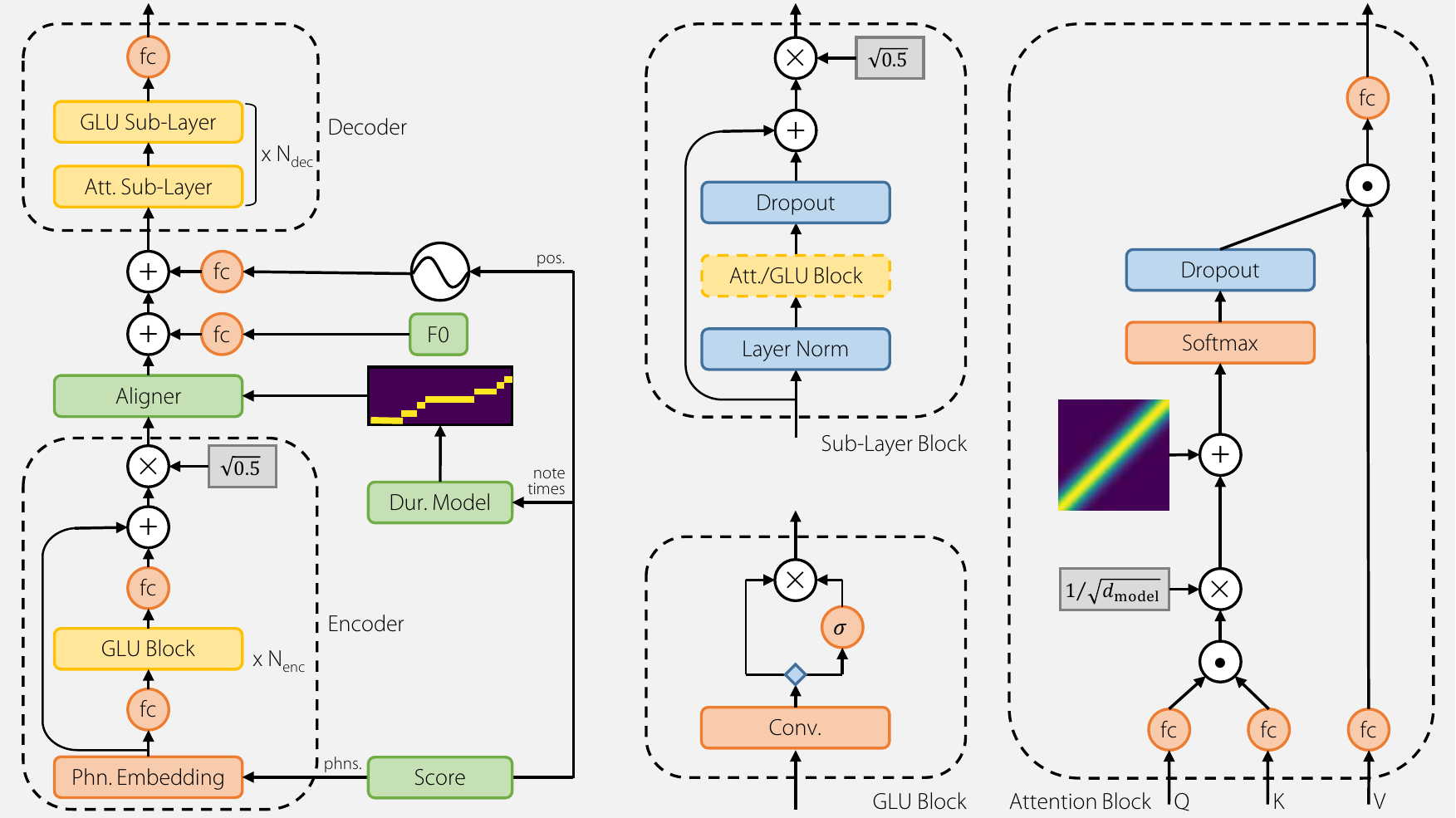}
  \caption{A diagram of the complete model architecture. On the left is the full system, composed of encoder, aligner and decoder, which themselves are composed of different higher level blocks. On the right, these higher level blocks (sub-layer, gated linear unit (GLU) and attention) are shown in detail.}
  \label{fig:model_architecture}
\end{figure*}

% components
The main components of our proposed system, as depicted in \cref{fig:model_architecture}, are the encoder, the aligner and the decoder. The encoder takes the input phonetic sequence and computes a sequence of hidden states corresponding to each phoneme and their local context. The aligner provides a hard alignment by repeating these states according to the predicted phoneme durations, obtaining a sequence of the same length as the output acoustic sequence. Next, some additional conditioning signals derived from F0 and position are added. The decoder, based on the Transformer model \cite{VaswaniA2017_AttentionIsAllYouNeed}, finally transforms the sequence of encoder hidden states to the target output sequence, through a series of self-attention and convolutional layers.

\subsection{Encoder}
Our encoder is based on the encoder proposed in \cite{PingW2018_DeepVoice3}. First, embeddings are computed from each input phoneme. Then, a series of convolutional blocks with gated linear units (GLUs) \cite{DauphinYN2017_GLU} allows encoding information about the phonetic context of each phoneme, e.g. corresponding to triphones or pentaphones. Finally a residual shortcut connection from the monophone embeddings is added to the local context output of the convolutional blocks.

\subsection{Duration model}
As noted, we purposely choose to use a very simplistic phoneme duration model in this work. It consists of a simple look-up table, populated with average phoneme durations computed from a dataset of a different singer with manually corrected phonetic segmentation. A simple heuristic is used to ensure that the sum of predicted phoneme durations matches the target note duration.

As we assume the note onset to correspond to the vowel onset, we first shift all onset consonants of each note to the preceding note (or silence). Then, we look up the sequence of average phoneme durations for each note, $[d_1,d_2,\dotsc,d_N]$, where $N$ is the corresponding number of phonemes, and $d_1$ corresponds to the duration of the vowel. In order to match the target note duration, $d_n$, we use the predicted consonant durations and fill the remaining duration with the vowel. However, we also ensure at least half of the note's duration is occupied by the vowel by fixing $r_v=0.5$. The scaling factor for the consonants, $r_c$, then becomes,
\begin{equation}
r_c = 
  \begin{cases}
  1                                                                    & \text{ for } N=1,\\
  \min\left(1, \cfrac{d_n - \rint{r_v d_n}}{\sum_{i=2}^{N} d_i}\right) & \text{ otherwise. }
  \end{cases}
\end{equation}
And, the adjusted phoneme durations, $[\hat{d}_1,\hat{d}_2,\dotsc,\hat{d}_N]$,
\begin{equation}
\hat{d}_i =
  \begin{cases}
  d_n - \sum\limits_{j=2}^{N} \max\left(1, \rint{r_c d_j}\right) & \text{ for } i=1,\\
  \max\left(1, \rint{r_c d_i}\right)                             & \text{ for } i=2,3,\dotsc,N.
  \end{cases}
\end{equation}
Note that all durations here are in integer number of frames, and that there are corrections for rounding errors and zero frame durations.

\subsection{F0 and position conditioning}
% F0: triangular
Continuous log F0 is encoded as a low dimensional vector between zero and one by evaluating several triangular basis functions whose centers are placed at frequencies appropriate for the training data's pitch range.

% position: cyclical
Transformers typically use an additive trigonometric positional encoding to give the self-attention blocks a sense of the position of their inputs, and provide a linear inductive bias along early on in training. However, in our case we found that a simple $K$-dimensional cyclical encoding of the normalized frame position within each note, $p \in [0,1] \subset \mathbb{R}$, gave slightly better results,
\begin{equation}
v = \frac{1}{2}\cos\left(2\pi p - 2\pi\frac{k-1}{K}\right) + \frac{1}{2}\quad\text{for }k=1 \ldots K.
\end{equation}

\subsection{Decoder}

Our decoder is based on a feed-forward variant of the Transformer model \cite{VaswaniA2017_AttentionIsAllYouNeed}, similar to \cite{RenY2019_FastSpeech}. Each layer consists of a self-attention sub-layer block and a convolutional sub-layer block. Both sub-layers blocks have layer normalization \cite{BaJ2016_LayerNorm}, dropout and a residual shortcut connection.

% self-attention blocks
Following \cite{VaswaniA2017_AttentionIsAllYouNeed}, our self-attention blocks use the scaled dot product as a scoring function. Additionally, similar to \cite{SperberM2018_GaussianBiasSelfAttention}, we bias the scores with a Gaussian along the diagonal to favor a more localized self-attention,
\begin{equation}
\Attention(Q,K,V) = \softmax\left(\frac{QK^{\transpose}}{\sqrt{d_{\text{model}}}} + M\right) V,
\end{equation}
\begin{equation}
M_{j,k} = \frac{-(j-k)^2}{2\sigma^2},
\end{equation}
where $d_{\text{model}}$ is the dimensionality of the input vectors, $M \in \mathbb{R}^{T \times T}$ for sequence length $T$, and $\sigma$ is a learned scale parameter. To reduce memory and computational requirements for the self-attention layers, we may use a reduction factor $r \geq 1$, which means $r$ frames are predicted per output timestep \cite{WangY2017_Tacotron,PingW2018_DeepVoice3}. While the use of multi-head attention is typical for NLP applications, we did not find this improved results in our case.

% convolutional blocks
For the convolutional blocks we use GLUs, which for our case outperform the 2-layer convolutional network with central ReLU activation typically used in Transformer architectures.

%%%%%%%%%%%%%%%%%%%%%%%%%%%%%%%%%%%%%%%%%%%%%%%%%%%%%%%%%%%%%%%%%%%%%%%%%%%%%%%%%%%%%%%%%%
\section{Experiments}\label{sec:experiments}
%%%%%%%%%%%%%%%%%%%%%%%%%%%%%%%%%%%%%%%%%%%%%%%%%%%%%%%%%%%%%%%%%%%%%%%%%%%%%%%%%%%%%%%%%%

% dataset(s)
For the experiments in this work, we train a model on a proprietary dataset of 41 pop songs performed by a professional English male singer. From this dataset 35 songs were used for training (\durhm{1}{26} total), 4 for validation and 2 for testing.

% hyperparameters
%- Feat dim: 60+4 ch
%- Hoptime: 10ms
%- Reduction factor: 2
%- Embedding dim: 256 ch
%- Embedding init: N(0, 0.01)
%- Encoder: 1x residual GLU 3x1 64 ch
%- F0 encoding: coarse 4 ch
%- Position in note encoding: cyclical 4 ch
%- Decoder: 6x FFTBlock with GLU 3x1, self-attention; 256ch
%- Output linear projection: 60*r
%- Dropout (encoder GLU blocks, decoder GLU blocks, decoder self-attention block, decoder sub-layer blocks): 0.9
%- Layer norm: sub-layer blocks
%- Self-attention Gaussian bias: learned scale, initialized to 30
%- Batch size: 32
%- Opt.: Adam, beta1=0.9, beta2=0.98, eps=1e-9, ema=0.995
%- LR: Noam (linear warmup, inverse square root decay) schedule with 4000 steps warm-up, 1e-3 base LR
%- Num. updates: 50k (1k epochs)
%- Init: Gehring
%- Loss: L1
Our proposed system uses 64-dimensional input features similar to \cite{BlaauwM2017_NPSS_MDPI}, extracted with a \SI{10}{\milli\second} hop time. A reduction factor, $r=2$, is used. We use 256-dimensional phoneme embeddings, and an encoder with a single 3x1 GLU block with 64 channels. F0 is coarse coded to a 4-dimensional vector, as is the position within the note, albeit with a cyclical encoding. The decoder consists of 6 layers with (single head) self-attention and 3x1 GLU blocks, all with 256 channels.  The final output projection is to $64r$ channels. Dropout probability is set to 0.1 throughout the model. The learned standard deviation of the Gaussian bias of the self-attention blocks is initialized to 30. Initialization of convolutional layers follows \cite{GehringJ2017_ConvAttention}. We use the Adam optimizer with $\beta_1=0.9$, $\beta_2=0.98$, $\epsilon=\num{1e-9}$, and a batch size of 32. We follow the learning rate schedule from \cite{VaswaniA2017_AttentionIsAllYouNeed}, with a 4000 step warm-up, a base learning rate of \num{1e-3}, and a total of \SI{50}{\kilo\nounit} updates. Additionally, we use Polyak averaging with a decay of 0.995 for validation and testing. The objective that we optimize is a simple L1 loss between output and target features.

% systems
We compare our proposed feed-forward model, which we label \mlabel{FFT-NPSS}, to an autoregressive baseline model roughly following \cite{BlaauwM2017_NPSS_MDPI}, labeled \mlabel{AR-NPSS}. To study the importance of the accuracy of the approximate initial alignment, we train a version of our model, which uses ground truth phonetic durations rather than predictions by the simple averages duration model. Note that the baseline \mlabel{AR-NPSS} is a non-Seq2Seq model, so it is also trained on ground truth phonetic durations. To study the importance of self-attention in the model we train a version of our model without attention blocks as well.

% listening test
We ran a MOS listening test with 18 participants, which each rated a random subset of 12 out of 20 phrases. Per test 6 stimuli were presented; the 4 systems mentioned previously, and visible and hidden references consisting of a WORLD re-synthesis of the target recording. All systems are presented and rated together to encourage a comparison between them.

\begin{table}
  \centering
  \caption{Mean Opinion Score (MOS) ratings on a 1--5 scale with their respective 95\% confidence intervals.}
  \label{tab:mos}
  \begin{tabular}[b]{@{}lcc@{}}
\toprule
\textbf{System}                        & \textbf{Mean Opinion Score} \\
\midrule
Hidden reference                       &         4.49 $\pm$ 0.09     \\
AR-NPSS                                &         2.56 $\pm$ 0.10     \\
FFT-NPSS (proposed)                    &         2.79 $\pm$ 0.11     \\
\textbf{FFT-NPSS w/ ground truth dur.} & \textbf{2.87 $\pm$ 0.11}    \\
FFT-NPSS w/o self-attention            &         2.48 $\pm$ 0.10     \\
\bottomrule
\end{tabular}

\end{table}

% results
The results of our listening test are shown in \cref{tab:mos}. We can see that the \mlabel{FFT-NPSS} system using ground truth phoneme durations performs best, but it is closely followed by the proposed Seq2Seq variant using a simple averages duration model. This shows that the initial alignment provided by the duration model has some importance, but it is not critical. Additionally, our proposed system outperforms the baseline autoregressive \mlabel{AR-NPSS} system, possibly due to avoiding issues related to exposure bias. Finally, the variant of the \mlabel{FFT-NPSS} system without self-attention layers performed worst overall, showing that self-attention is an important component for this kind of Seq2Seq mechanism, in our observations especially in terms of providing a coherent timbre over time. While all systems are still rated considerably below the reference WORLD re-synthesis, we expect that this gap would be reduced if we combine our system with a neural vocoder. Some sound examples, both with and without neural vocoder, are available online\footnote{\url{https://mtg.github.io/singing-synthesis-demos/transformer/}}.

%%%%%%%%%%%%%%%%%%%%%%%%%%%%%%%%%%%%%%%%%%%%%%%%%%%%%%%%%%%%%%%%%%%%%%%%%%%%%%%%%%%%%%%%%%
\section{Relation to prior work}\label{sec:prior_work}
%%%%%%%%%%%%%%%%%%%%%%%%%%%%%%%%%%%%%%%%%%%%%%%%%%%%%%%%%%%%%%%%%%%%%%%%%%%%%%%%%%%%%%%%%%

% FastSpeech
Our work is most closely related to the recently proposed FastSpeech model for TTS \cite{RenY2019_FastSpeech}. This model is also based on the feed-forward Transformer and an initial alignment obtained from a duration model. However, in this case the duration model is trained with the help of a teacher model based on an autogressive Transformer \cite{LiN2019_TransformerTTS}, which is also used for generating the target mel-spectrogram features. We wanted to avoid the need to train an autoregressive teacher model, as we found this generally challenging for the case of singing voice. Additionally, we apply some modifications to the architecture, such as the use of GLU convolutional blocks, alternative positional encoding and a Gaussian bias for the self-attention layers.

% ParaNet
The ParaNet model \cite{PengK2019_ParaNet} proposes a different approach to feed-forward TTS. Here, standard content-based encoder-decoder attention is used, but the model is trained trained with the help of attention distillation with an autoregressive teacher model based on \cite{PingW2018_DeepVoice3}. Besides the reasons mentioned above, we found that the hard alignment used in our approach makes it easier to obtain a quality similar to non-Seq2Seq models, compared to the soft alignment of encoder-decoder attention.

% Korean Singing Synthesizer
In singing synthesis, the only Seq2Seq system we are aware of is \cite{LeeJ2019_KoreanSS}. This model is based on the DCTTS model \cite{TachibanaH2018_DCTTS}, using content-based encoder-decoder attention, with autoregressive decoder. Similar to our approach, there is an initial alignment of the input states to the output timesteps. However, relying on the fact that the Korean syllable structure has at most one onset and one coda consonant, the first and last frame of the note are assigned to each consonant respectively, and the remaining frames are assigned to the vowel. After which, learning the attention alignment can be facilitated by using diagonally guided attention \cite{TachibanaH2018_DCTTS}.

% Non-seq2seq singing synthesis systems
% AR: NPSS, IRCAM, DAR
% CNN down/up: TechnoSpeech
% GAN: WGANSing, Nitech
Non-Seq2Seq singing synthesizers include those based on autoregressive models \cite{BlaauwM2017_NPSS_MDPI,BousF2019_SpecEnvSingSynth,YiYH2019_DAR_SS}, feed-forward CNN \cite{NakamuraK2019_TechnoSpeechCNN}, and feed-forward GAN-based approaches \cite{HonoY2019_GAN_SS,ChandnaP2019_WGANSing}.

%%%%%%%%%%%%%%%%%%%%%%%%%%%%%%%%%%%%%%%%%%%%%%%%%%%%%%%%%%%%%%%%%%%%%%%%%%%%%%%%%%%%%%%%%%
\section{Conclusions}\label{sec:conclusions}
%%%%%%%%%%%%%%%%%%%%%%%%%%%%%%%%%%%%%%%%%%%%%%%%%%%%%%%%%%%%%%%%%%%%%%%%%%%%%%%%%%%%%%%%%%

% conclusions:
% - presented Transformer-based model with simple seq2seq approach
% - as good or slightly better than ar baseline (and faster, no exposure bias issues)
% - self-attention critical
% - accurate duration model only slightly important; and observed model can recover from errors
We presented a singing synthesizer based on the Transformer model, with a practical Seq2Seq mechanism allowing feed-forward operation. This approach allows training models without the need for pre-aligned training data, which can be cumbersome to prepare for singing data. Compared to a baseline autoregressive model, the proposed model allows for faster inference, avoids issues related to exposure bias, and rates as good or slightly better in listening tests. The use of self-attention resulted to be a key factor in obtaining good quality results, especially in terms of producing coherent timbre. As our model relies on an initial alignment provided by a duration model, we compared a very simplistic duration model to ground truth durations, to see the importance of the initial alignment's accuracy. In listening tests, using ground truth durations was rated highest, but the difference was relatively small. While not shown due to lack of space, in our observations the model can recover from errors in the initial alignment, most likely thanks to the decoder's non-causal convolutions and self-attention layers. For example, while the duration of phrase-final consonants tends to be systematically underpredicted by average durations, in the output of the synthesizer these phonemes have durations close to the target.\thanksanywhere{This work was funded by TROMPA H2020 No 770376.}

% future work:
% - more effective regularization (powerful model; currently has quite big generalization gap)
% - F0 modeling with similar seq2seq approach, and architecture
% - training from symbolic note timings rather than vowel onsets
% - combination with feed-forward neural vocoder (ar vocoders seem a little more robust to oversmoothing in conditioning features)

%%%%%%%%%%%%%%%%%%%%%%%%%%%%%%%%%%%%%%%%%%%%%%%%%%%%%%%%%%%%%%%%%%%%%%%%%%%%%%%%%%%%%%%%%%
% References
%%%%%%%%%%%%%%%%%%%%%%%%%%%%%%%%%%%%%%%%%%%%%%%%%%%%%%%%%%%%%%%%%%%%%%%%%%%%%%%%%%%%%%%%%%
%\vfill\pagebreak
\clearpage

%\section{References}
%\printbibliography[heading=none]  % !LuaLaTeX

\bibliographystyle{IEEEbib}
{\small\bibliography{transformer}}

\end{document}